\documentclass[structabstract]{aa}  
\usepackage{graphicx}
\usepackage{txfonts}

\newcommand{\ho}{{\rm km~s}^{-1}~{\rm Mpc}^{-1}}
\begin{document}
   \title{Deep near-infrared imaging of the HE0450-2958 system }
   \author{G. Letawe
          \inst{1}
          \and
          P. Magain\inst{1}
          }

   \institute{ $^1$ Institut d'Astrophysique et de G\'eophysique, Universit\'e de Li\`ege, all\'ee du 6 ao\^ut 17, B-4000 Li\`ege\\
              \email{gletawe@ulg.ac.be}
             \thanks{based on observations made at ESO La Silla/Paranal observatory, Chile, program 082.B-0288(B)}
             }

   \date{Received; accepted -}
 \authorrunning{G. Letawe \& P. Magain}
\titlerunning{HE0450-2958: deep NIR imaging}
 
  \abstract
   {The QSO HE0450-2958 and the companion galaxy with which it is interacting, both ultra luminous in the infrared, have been the subject of much attention in recent years, as the quasar host galaxy remained undetected.  This led to various interpretations on QSO and galaxy formation and co-evolution, such as black hole ejection, jet induced star formation, dust obscured galaxy,  or  normal host below the detection limit. }
   {  We carried out deep observations in the near-IR in order  to solve the puzzle concerning the existence of any host.}
   { The object was observed with the ESO VLT and HAWK-I in the near-IR  J-band for 8 hours. The images have been processed with the MCS deconvolution method (Magain, Courbin \& Sohy, 1998), permitting accurate subtraction of the
QSO light from the observations.}
   {The compact emission region situated close to the QSO, called the blob, which  previously showed only gas emission lines in the optical spectra, is now detected in our near-IR images.  Its high brightness implies that stars likely contribute to the near-IR emission.  The blob might thus be interpreted as an off-centre, bright and very compact host galaxy, involved in a violent collision with its companion.  }
   {}

   \keywords{Quasar: individual: HE0450-2958
               }

   \maketitle
%

\section{Introduction}


 While there is convergence in the scientific community on the idea that the super massive black holes (SMBHs) lying at the centre of galaxies are growing in parallel with the galactic bulge they are hosted in, a few outstanding objects seem to lie outside this relation.
This is probably the case of HE0450-2958.  Located at z=0.286, 04h 52m 30.10s $-$29d 53m 35.3s, this system was first detected as an IRAS source (\cite{degrijp}), then identified as a QSO and studied as an interacting active galaxy (\cite{low}, \cite{canalizo}). The interest in this system grew with the claim by Magain et al.\  (2005) that no host galaxy was detected on optical HST images (ACS F606W), implying that it was either abnormally compact  or more than six times fainter than  the QSO hosts  described in \cite{floyd04}.   The only significant extended emission arose from two objects in the nearby field:  first, from a strongly disturbed companion galaxy 7\,kpc distant in projection from HE0450-2958, and also from a cloud of highly ionized gas showing no trace of stellar light, centered at 0.9\,kpc NW from the QSO.

Several studies followed, proposing various scenarios for explaining this system. \cite{haenelt} and \cite{hoffman07} proposed that it was the first observation of a BH ejected during a galactic merger. \cite{merrit} observed that the QSO spectrum was similar to those of Narrow Line Seyfert 1 galaxies, suggesting accretion above the Eddington limit and thus argued that the BH mass had been overestimated and that the host could then be just below the detection limit. Radio observations of the CO  molecule by \cite{papad} implied that the bulk of star formation was located in the companion galaxy  and not around the QSO, and that both the QSO and galaxy have the properties of Ultra Luminous Infrared Galaxies (ULIRGs). \cite{letg09} and \cite{knud09} showed that the center of the companion galaxy is highly obscured by dust. \cite{letg09} also presented evidence that a second AGN lies in the companion galaxy, hidden by the thick dust cloud.  Lately \cite{elbaz} proposed an alternative scenario of galactic building, by jet induced star formation.

None of these studies has however been able to  directly detect a host galaxy.  This is the purpose of the present deep near-IR observations with the ESO VLT, aimed at detecting a co-centered host galaxy, if any, around the QSO.
\begin{figure}
 \centering
\includegraphics[width=9cm]{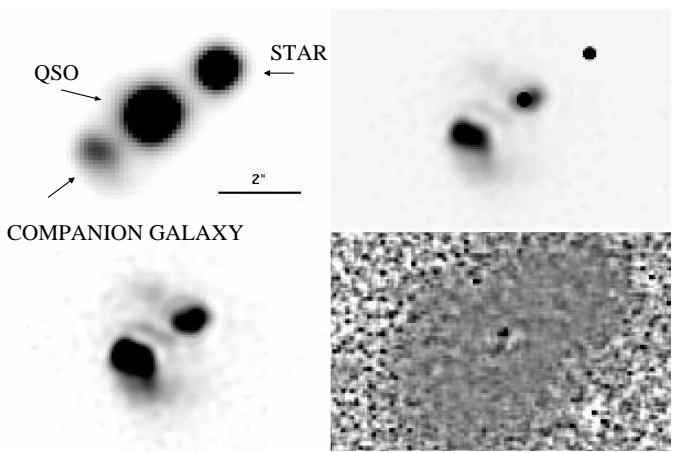}
\caption{HAWK-I observation of HE0450-2958 in the J-band. Top left: one of the 12 exposures; Top right: deconvolved image, where the point sources are  modeled by 2D gaussians, here appearing as black dots; Bottom left: diffuse background only (companion galaxy + other extended emissions)  as obtained by simultaneous deconvolution of all exposures (see text); Bottom right: average residuals  (model minus observations, in units of the standard deviation; the intensity scale goes from $-3 \sigma$ to $3 \sigma$) . North is up and East to the left.}
\label{imgdec}
\end{figure}


\section{Observations }

The QSO HE0450-2958 and its surroundings were observed with HAWK-I (\cite{hawk1,hawk2,hawk3}), mounted on  Unit Telescope 4 of the ESO/VLT at Paranal Observatory,  for 8 hours in the near-IR J-band. The observations were  divided into 12 exposures of 40 min each, obtained between October 2008 and January 2009, with  variable seeing ranging from 0.6 to 1.0\arcsec (0.8\arcsec on average). One exposure was unusable because of technical problems.  The observations were automatically processed with the ESO Pipeline. The field of view is 7.5\arcmin $\times$ 7.5\arcmin\  wide, with a spatial resolution of 0.106\arcsec/pix. As no standard star was observed  and as the Hawk-I J-band is identical to  the  2MASS J-band, we used four 2MASS stars (\cite{2mass}), with magnitudes from J=12.9 to 14.6, to compute the photometric zero-point.  The four stars give essentially the same result, with a dispersion below 0.02 mag.  The depth of the observations allows us to detect isolated objects down to J=25,  at 5 $\sigma$. Conversions to absolute magnitudes were made using the following cosmological parameters: $H_0=71~\ho $, $\Omega_m=0.3$ and $\Omega_{\Lambda}=0.7$, leading to a scale of 4.247 kpc/\arcsec\, and a luminosity distance of 1450 Mpc at the redshift of the object.

\section{Image processing and results}
 To analyze these images we used the deconvolution method described in Magain, Courbin, \& Sohy (1998; hereafter MCS).  MCS is a powerful tool for QSO host analysis, as it  permits efficient separation of point sources from diffuse components. For optimal results, it requires a very accurate Point Spread Function (PSF). For this purpose, we built our PSF by using the foreground star close to the system, in order to minimize the effects of PSF variation across the field. We  eliminated the slight contamination of the PSF star by the QSO light, using the iterative method described in \cite{chantry}.  One advantage of the MCS method is that it  allows simultaneous treatment of all exposures, resulting in a single diffuse background component common to all the 11 exposures, and in precise astrometry and photometry for the point sources. The decomposed and deconvolved images are presented on Fig.~\ref{imgdec}. They show a compact, but clearly extended, emission close to the QSO.
\begin{figure}[h!!]
 \centering
\includegraphics[width=9.cm]{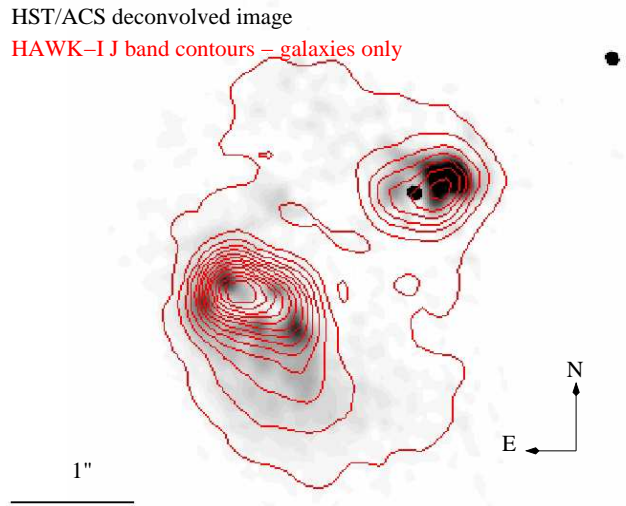}
\caption{ The deconvolved HST/ACS F606W image (\cite{mag05}) of the HE0450-2958 system is shown in greyscale.  Red contours correspond to the deconvolved HAWK-I $J$-band image, after removal of point sources.}
\label{contour}
\end{figure}
 
As the QSO is significantly brighter than the PSF star, we carefully checked if the data were not suffering from  nonlinearity, even if the observations are all well below the saturation limit.  Indeed, a departure from linearity at the approach of saturation would imply a flatter PSF for more intense sources (e.g. the QSO), and the deconvolution process might compensate this effect by adding a wider background component. We ran separate deconvolutions on subgroups of data, sorted by seeing, the  nonlinearity being expected to be stronger for better seeing, i.e. higher peak intensity.  No significant variation of the background intensity is found between the different subgroups, implying that if any deviation from linearity is present, it has a negligible effect on the results.  

The deconvolved images  reveal a bright extended but very compact component near the QSO. Residuals are not perfectly satisfactory at the position of the QSO, but their amplitudes are really small compared to the detected emission (the largest residual is about 10\% of the peak in the extended emission and 0.1\% of the QSO intensity).

The observed emission is surprisingly compact, with a measured effective ( half-light) radius of only 1 kpc, which is about the size of the most compact elliptical galaxies.  Moreover, as this effective radius is smaller than the original resolution of the observations, it should be considered as an upper limit. On Fig.~\ref{contour} we show the HAWK-I J-band contours of the extended emissions, after removal of the point sources, overlaid on the HST/ACS F606W-band deconvolved image of the system  (from \cite{mag05}). 
It shows that the compact emission nicely matches the ``blob'' of ionized gas discovered in the HST/ACS images, in which only optical emission lines  and no stellar continuum were detected spectroscopically, see Magain et al.\ (2005), \cite{letg08}. The measured projected distance between the QSO and the peak emission in the extended component also agree in the two spectral bands, with values of 0.218\arcsec (0.93 kpc) in the HST/ACS image and 0.212\arcsec (0.90 kpc) in the VLT/HAWK-I image.  We can thus safely associate the emission detected in the near-IR with the blob discovered in the optical HST image.  

\begin{figure}
 \centering
\includegraphics[width=8.cm]{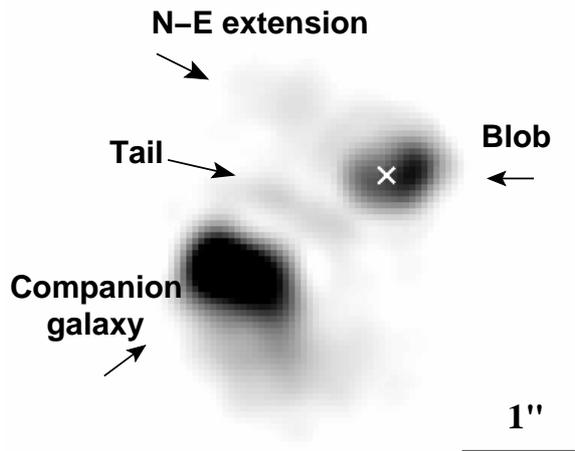}
\caption{Zoom on HE0450-2958 system, with point sources removed (the QSO position is indicated by the white cross). The different components are indicated: companion galaxy, blob, N-E extension as in \cite{knud09}, and the newly discovered 'tail' feature.}
\label{zoom}
\end{figure}

Additional emissions are also detected: the N-E extension previously reported in \cite{knud09}, and a kind of elongated emission region (that we call {\it tail})  between the QSO and the companion galaxy. Both these emission regions seem to be disconnected from the blob.  Photometry of the different components is reported in Table \ref{mags}, with associated locations found in Fig.\ref{zoom}.

\begin{table}
\centering
\caption{Photometry of the different components. Apparent magnitudes are given in the Vega based J-band.  Magnitudes derived from the different exposures are very stable, resulting in $1 \sigma$ uncertainties below 0.02 mag.}
\begin{tabular}{|c|c|cccc|c}
\hline  Global & QSO & Blob & NE & Tail & Companion \\ 
system&only& & ext.&  & galaxy \\
\hline 
13.74& 14.28&17.68&19.15&18.92&16.42\\  
\hline
\end{tabular}
\label{mags}
\end{table}

\section{Discussion}

\subsection{The blob}

Given the clear near-IR detection of the blob, one of the questions which naturally arises is whether we have detected a stellar component, which would mean that the so-called blob could be considered as a galaxy (i.e. a collection of stars), even if quite an atypical one.

As we have no NIR spectrum of the blob, we can only give a rough estimate of the contribution of the gas emission lines to the overall flux in the J-band.  Given the redshift of the blob, the AGN emission line which dominates by far in this spectral region is the [SIII] line at 953 nm.  On the other hand, the most intense emission line of the blob spectrum in the F606W spectral band is the [OIII] line at 501 nm.  From typical AGN spectra (\cite{osterb}), the ratio of the [OIII] to [SIII] lines amounts to at least a factor 3 (we recall that the optical line ratios of the blob are typical of gas excited and ionized by AGN radiation, \cite{mag05}).

Using the optical VLT spectrum of the blob, we compute that the [OIII] line contributes in between 18\% and 40\% of the flux in the F606W image, depending on how well the blob spectrum could be separated from the nearby and much brighter QSO spectrum.  The apparent magnitude of the blob in the F606W HST/ACS image is 19.57, while it amounts to 17.68 in the J-band HAWK-I image.  Converting apparent magnitudes in fluxes and assuming that the [SIII] line flux is at most one third of the [OIII] flux, we compute that it would contribute at most 2 to 5\% of the measured flux in the J-band.  It is thus likely that we need a continuum contribution to explain the observed J-band flux.  We can thus conclude that the blob most probably contains stars.  In this context, the non-detection of the stellar continuum in the VLT spectrum of the blob (\cite{mag05}) can be explained by the fact that it lies below the detection limit in this 30 min spectrum, which is to be compared with 8 hours of exposure in the present broad-band image.

Let us thus assume that most of the J-band flux of the blob is due to stars.  Its average surface brightness measured within the effective radius amounts to $ < \mu_e^J>= 16.6$,  matching  the compact and high surface brightness end of the elliptical domain of the Kormendy relation in the near-IR, see for instance \cite{pahre}.  
Such an extreme surface brightness can be found in the most compact elliptical or S0 galaxies and lies at the limit of the brightest spiral galaxies (\cite {mcdo}, \cite{mollenhof}).  The surface brightness of the blob is thus high,  but not outside the range
observed for extremely compact objects.

Let us now estimate, if the blob (despite its $\sim$1kpc offset from the QSO) were to be interpreted as the bulge of the host galaxy, where it would lie in the BH-bulge relation. The observed J-band magnitude of the blob (J=17.68) converts to absolute M$_{\rm{Jobs}}=-23.4$. 
Following the broad band colours given in \cite{buzzoni05}, this magnitude translates to restframe M$_{\rm{J}}=-24.0$ for an old  stellar population in a bulge or elliptical (it would only change to M$_{\rm{J}}=-23.8$ for a young stellar population in an Im galaxy). 
The relation between bulge luminosity and BH mass given in \cite{marconi} thus predicts a BH mass of 4.7$\times$10$^8$ M$_{\odot}$, between five and ten times larger than the previous estimated BH masses. Indeed, as reviewed in \cite{knud09}, these estimates range between 4 and 9$\times$10$^7$ M$_{\odot}$.  Thus either the blob is not only composed of a bulge, or it does not lie on the usual BH-to-bulge relation (note, however, that, given the 0.3 dex scatter of the \cite{marconi} relation, the deviation from the mean would only amount to 2.4 to 3.6$\sigma$).    In conclusion, the observed magnitude and offset position give evidence that the blob cannot be considered as a host bulge. 

\subsection{Detached emissions}
 As the position of the `tail' is aligned with the radio jet direction (\cite{feain}), jet induced star formation as proposed by \cite{elbaz} might have formed this structure, pushing it towards the companion galaxy. Previous spectral analysis of the system in \cite{lety08} had already mentioned some gas reflecting the quasar radiation and moving away from the QSO towards the companion galaxy. This gas finds here a likely stellar counterpart in the `tail',  consistent with the hypothesis of material blown by the quasar radiation and where star formation is activated by the radio jet. 

This scenario however does not explain the creation of the N-E extension, located outside the bulk of star formation in the companion galaxy and not really connected to the blob.
If the jet induced star formation scenario is not suited because of misalignement, it is to notice that a simpler and less exotic scenario, i.e. a violent collision between a gas rich galaxy (the companion) and another galaxy, likely associated with the bright QSO, might explain the different extended emissions as parts of the galaxies disrupted by the gravitational interaction.



\subsection{Status of the second AGN}
We previously reported in \cite{letg09} the possible detection of a second active nucleus in the companion galaxy, hidden by large dust clouds.  This tentative detection was based on HST/NICMOS H-band and VLT/ISAAC K-band observations.

We looked for a point source in the J-band image of the companion galaxy and could not detect any. This is however compatible with an AGN hidden behind a dust cloud, the extinction in the J-band  being sufficient to make this point source disappear almost completely ( the estimated  extinction $A_V \simeq 15$  converts to $A_I \simeq 7.2$,  as the observed J-band corresponds roughly to restframe I-band).

If confirmed, the presence of a second AGN in the companion galaxy would be difficult to explain if the latter was built up only by star formation induced by a jet originating from the main QSO, as suggested in \cite{elbaz}.

\section{Conclusion}
The  detection of a plausible host rules out the  scenario of the ejection of a BH from the companion galaxy, that was previously introduced by \cite{haenelt} or \cite{hoffman07} and discussed in several following papers concerning HE0450-2958, such as \cite{merrit} and \cite{letg09}. If the  BH had been ejected with some gas to feed it, it would not be accompanied by a galaxy with stars.

  The well-detected stellar emission from the blob sheds new light on recent attempts to explain this unusual system. 
Indeed, \cite{elbaz} proposed that the activity of the quasar and associated radio jets have been creating the companion galaxy and parts of the host in construction, still partially disjoint from the QSO as seen in a primordial step of its evolution.  The observation of stars in the direct vicinity of the QSO shows that there is already a host candidate, independently of the impact of the radio jet on the surrounding objects. Moreover, from the CO observations of \cite{papad}, no star formation is found at the blob location, excluding the possibility that the QSO could have created the blob.

 These findings support the hypothesis already presented by \cite{papad}, involving a collision between a gas rich galaxy (the companion), and a smaller one (the blob). Two galaxies have collided, each of them probably harbouring an active BH, the interaction enhancing the star formation in at least one of the galaxies and disrupting the pre-existing structures.  The disrupted parts will probably merge in the future into a single galaxy. The radio jet would play a secondary role in the enhancement of the star formation in some parts of the system (e.g.\ in the ``tail" in between the main QSO and the companion galaxy, and maybe in the N-E part of the companion galaxy itself).  Moreover, the QSO radiation would ionize the gas expelled all around the system (Letawe et al., 2008b)  

 In fact, this scenario  explains all the observations: disturbed morphology in both galaxies (including off-centre nuclei), presence of the second AGN, enhanced star formation, N-E stellar emission and extended emission line regions. The previous non detection of the stellar content in the blob was due to its faintness in the optical, its compactness and its proximity to the bright QSO, bringing the optical continuum just below the detection limits. The secure detection of stars near the QSO, even if  not centred on the nucleus, makes the HE0450-2958 system  consistent with a highly perturbed merger, probably not requiring more exotic scenarios. As such, it fits reasonably well with the hierarchical models of galactic building by successive mergers.

\begin{acknowledgements}
GL is a teaching assistant supported by the University of Li\`ege. This work was also supported by PRODEX Experiment Agreement 90312 (ESA and PPS Science Policy, Belgium).
\end{acknowledgements}

\end{document}